\documentclass[usenatbib]{mn2e}
\usepackage{psfig}

\title{Radio Observations of the Supernova Remnant Candidate G312.5-3.0}
\author[S.R. Kane and A.E. Vaughan]{Stephen R. Kane$^1$ and Alan E.
Vaughan$^2$\\
$^1$School of Physics \& Astronomy, University of St Andrews, North Haugh,
St Andrews, Fife KY16 9SS, Scotland\\
$^2$Department of Physics, Macquarie University, Sydney, NSW 2109,
Australia}

\begin{document}

\maketitle

\begin{abstract}
The radio images from the Parkes-MIT-NRAO (PMN) Southern Sky Survey at
4850 MHz have revealed a number of previously unknown radio sources. One
such source, G312.5-3.0 (PMN J1421-6415), has been observed using the
multi-frequency capabilities of the Australia Telescope Compact Array
(ATCA) at frequencies of 1380 MHz and 2378 MHz. Further observations of
the source were made using the Molonglo Observatory Synthesis Telescope
(MOST) at a frequency of 843 MHz. The source has an angular size of
$18\arcmin$ and has a distinct shell structure. We present the reduced
multi-frequency observations of this source and provide a brief argument
for its possible identification as a supernova remnant.

\end{abstract}


\section{Introduction}

Extensive radio surveys of the sky began in the 1950s when the number of
identified sources was not sufficient to conduct significant statistical
studies. The large-area continuum radio surveys have become the foundation
for all other radio astronomy tests since they provide the necessary radio
source catalogues and the location of objects with which astronomers can
calibrate their instruments.

The Parkes-MIT-NRAO (PMN) survey of the southern sky was undertaken as a
collaboration between the Parkes Radio Observatory of the Australia
Telescope National Facility (ATNF), the Massachusetts Institute of
Technology (MIT), and the National Radio Astronomy Observatory (NRAO)
\citep{gri93}. The primary aim of the project was to survey the whole
southern sky at a frequency of 4850 MHz in order to provide a more
extensive, high-quality catalogue of radio sources for the Southern
Hemisphere. The secondary aim was to use this survey to complement the
results obtained from the Northern Hemisphere 4850 MHz survey performed by
\citet*{con89} which was made using the NRAO 91 m telescope \citep{wri94}.
Previously, the only large-area, high-frequency radio survey in the same
area was the 2700 MHz survey of Bolton and his collaborators which
contains around 8200 sources \citep{gri91}.

The Southern Survey $(-87.5\degr<\delta<-37\degr)$ and the Tropical Survey
$(-29\degr<\delta<-9.5\degr)$ increased the number of known radio sources
in the Southern and Tropical zones by a factor of about 5 over those known
previously from the Parkes 2700 MHz and Molonglo 408 MHz surveys
\citep{gri93}. The catalogue of radio sources discovered in the Southern
Survey contains 23,277 sources. However, it is important to realise that
the PMN surveys primarily detected point sources and the reduction process
was optimised for such objects. This resulted in the exclusion of many
extended objects from the survey with the exception of stronger sources
which had a sufficiently high signal-to-noise ratio \citep{wri94}.

\begin{figure}
\psfig{file=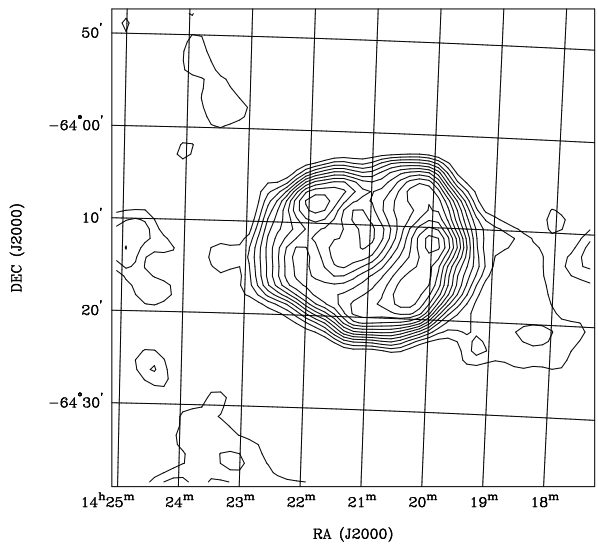,width=7.8cm}
\caption{A contour map from the PMN survey showing the object of interest.
The contour levels start at 10 mJy/Beam and increase in steps of
10 mJy/Beam.}
\end{figure}

It was from an image of the PMN survey that the chosen object of study was
discovered. The object is at right ascension $14^{\mathrm{h}}20^{\mathrm{m}}$
and declination $-64.2\degr$ (Galactic longitude $312.5\degr$ and latitude
$-3\degr$) and has an approximate diameter of $18\arcmin$. The object is
shown towards the centre of Figure~1 which is a map taken from the PMN
survey. The circular appearance of the object suggests it may be a supernova
remnant (SNR) although it does not currently appear in SNR catalogues. The
source has also been noted as a SNR candidate by \citet{dun97}. The object
seems to have no optically visible counterpart and no counterpart has been
found in current ROSAT data. Shown in Figure 2 is an IRAS 12 micron image
from the same region of sky as the object \citep{mcg96}. This indicates that
the object may have a strong counterpart in the infra-red, although dust in
the local environment makes further conclusions based on the IR data
difficult.

\begin{figure}
\psfig{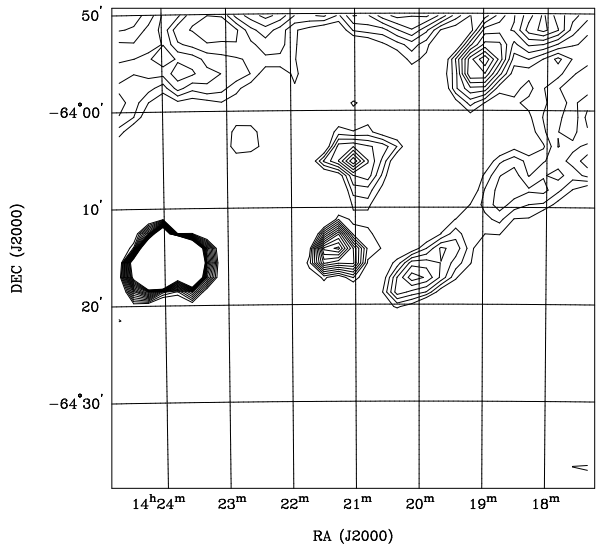}
\caption{A contour map from the IRAS 12 micron survey showing the same
region of sky. The contour levels start at 380 mJy/$\sq\arcmin$ and
increase in steps of 8 mJy/$\sq\arcmin$.}
\end{figure}

\section{Observations and Data Reduction}

The extended source was observed in January, 1995 using the mosaic
capabilities of the Australia Telescope Compact Array (ATCA). The
observation used a 19 pointing centre mosaic in a hexagonal grid with the
pointing centres set $12\arcmin$ apart from each other. The radio sources
1934-638 and 0823-500 were used as the primary calibrators and 1549-790
was used as the secondary calibrator. The object was observed at
frequencies of 1380 MHz and 2378 MHz which, along with a baseline of 375 m,
were chosen to approximately match the resolution of the PMN survey at
4850 MHz. Each band had a width of 128 MHz divided into 32 channels.

The allocated observing time began at 10:30 UT, January 17 and extended
through to 2:30 UT, January 18 and allowed an approximate 12 hour
synthesis observation of the source. The observations are summarised in
Table 1.

\begin{table}
\caption{ATCA observations of G312.5-3.0}
\begin{tabular}{@{}lcc}
Parameter & 1380 MHz & 2378 MHz\\
Configuration (m) & 375 & 375\\
Bandwidth (MHz) & 128 & 128\\
Integration time (min) & 720 & 720\\
Beam Size (FWHM) & 115.99\arcsec $\times$ 128.83\arcsec & 67.31\arcsec
$\times$ 74.76\arcsec\\
rms noise (mJy beam$^{-1}$) & 0.029 & 0.040\\
Diameter & 20\arcmin & 20\arcmin\\
\end{tabular}
\end{table}

The reduction of the ATCA data was performed using the software package
Miriad which was developed by the Berkeley Illinois Maryland Association
(BIMA) and then later adapted to the needs of the ATNF \citep*{sau95}. The
software package contains tasks designed to reduce data from mosaic
observations. The results from these reductions are presented in the next
section.

Further observations of the source were made using the Molonglo
Observatory Synthesis Telescope (MOST) at a frequency of 843 MHz which
served to complement the data obtained from the Compact Array. The
observation was carried out in September, 1995 and the observed field has
an area of $70\arcmin \times 70\arcmin$. The phase gradient of the MOST,
together with the tilting of the east-west arm about its long axis,
provides the equivalent of an alt-alt mounting which has full uv-coverage
for declinations south of $-30\degr$ \citep{mil81,mil85}.

\section{Results and Discussion}

Radio images of the source have been produced from the observations. The
images shown in Figures 3, 4, and 5 are the contour maps of the extended
source at frequencies of 843 MHz, 1380 MHz, and 2378 MHz respectively.
These images may be used as a first step to determine some of the
astrophysical properties of the source in an attempt to discover its
nature and the emission processes which are occurring. This information
assists in constraining the identity of the source as either a H I region,
a H II region, a SNR, or a planetary nebula.

\begin{figure}
\psfig{file=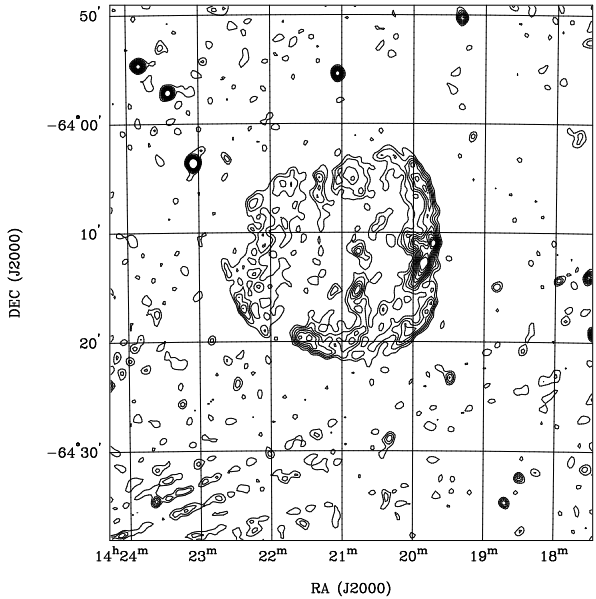,width=7.8cm}
\caption{A contour map of the extended source at 843 MHz. The contour
levels start at 2 mJy/Beam and increase in steps of 2 mJy/Beam.
Structure in the image due to a nearby strong source just outside the
field is also visible.}
\end{figure}

\begin{figure}
\psfig{file=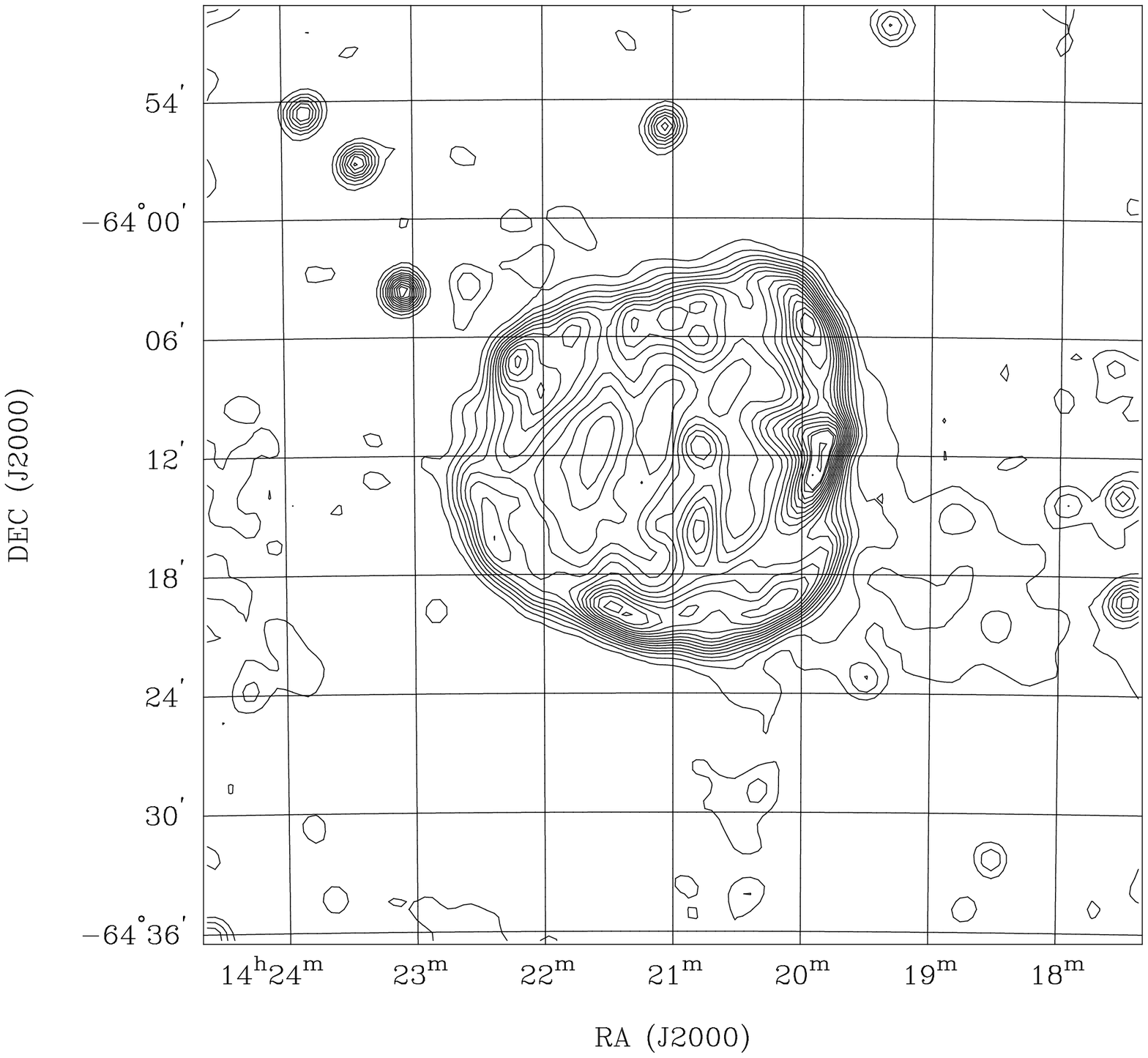,width=7.8cm}
\caption{A contour map of the extended source at 1380 MHz. The contour
levels start at 2 mJy/Beam and increase in steps of 2 mJy/Beam.}
\end{figure}

\begin{figure}
\psfig{file=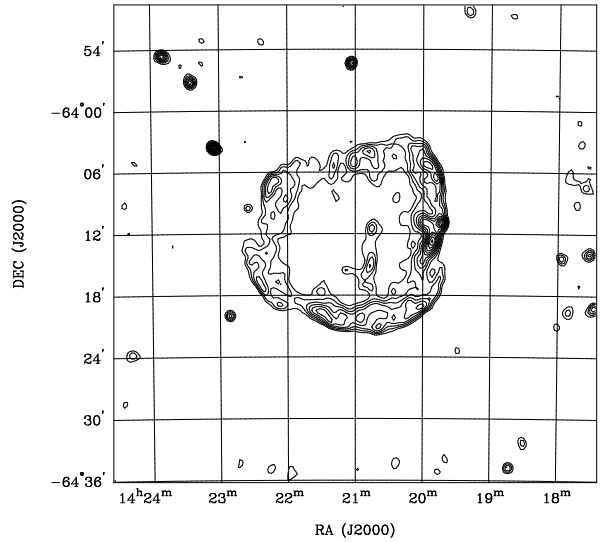,width=7.8cm}
\caption{A contour map of the extended source at 2378 MHz. The contour
levels start at 2 mJy/Beam and increase in steps of 2 mJy/Beam.}
\end{figure}

\subsection{Morphology}

At each frequency, the object appears to be faint in the middle with a
ring structure around the edge. This indicates that the object is likely
to be a spherical shell, which is a structure typical of planetary nebulae
and shell remnants produced by Type I supernovae
\citep{min79,min85,mal91,all94}. The symmetry of the images also suggests
that the extended region originated at a central location whereas H I and
H II regions tend to be amorphous clouds with no axis of symmetry. At the
side of smaller right ascension, the intensity of the object appears to
be higher, indicating that this region may have more mass or is more dense
(or both). This is probably due to interactions with the interstellar
medium. The observed shell structure suggests that the source is optically
thin at the observed frequencies.

For an optically thin source, the equation of radiative transfer can be
approximated by
\begin{equation}
I_\nu = S_\nu \tau_\nu
\end{equation}
where the source function $S_\nu$ is defined as $S_\nu \equiv j_\nu /
\alpha_\nu$ and where $\alpha_\nu$ and $j_\nu$ are the absorption and
emission coefficients of the medium respectively.

Assuming the source function to be constant throughout the source,
equation (1) may be expressed as a ratio
\begin{displaymath}
\frac{I_1}{I_2} = \frac{\tau_1}{\tau_2}
\end{displaymath}
where the subscripts refer to two paths through the source. By measuring
the flux density at the edge and at the middle of the object, this ratio
was estimated to be $I_1 / I_2 = 2.27$.

The absorption coefficient will be constant throughout the source so
from the definition of optical depth
\begin{displaymath}
\frac{\tau_1}{\tau_2} = \frac{s_1}{s_2}
\end{displaymath}
where $s$ is the path length. By estimating the width of the shell to
be approximately $3\arcmin$, the path lengths through the middle and
through the edge were calculated. The ratio was then estimated to be
$s_1 / s_2 = 2.23$. The near equality of the two ratios confirms that the
source is optically thin, based on the assumption of constant absorption
within the source.

\subsection{Flux Densities}

The estimated total flux density at each frequency is shown in Table 2.
The errors associated with the flux density measurements at each frequency
are mainly due to (unknown) losses of flux density in the lower spatial
frequencies for each of the images. Due to the low-pass filter applied to
the PMN data during the reduction process, it should be noted that the flux
given at 4850 MHz is likely a lower limit to the flux at this frequency.

\begin{table}
\caption{Total flux density at each frequency.}
\begin{tabular}{@{}cc}
Frequency (MHz) & Flux Density (Jy)\\
843 & $1.1 \pm 0.2$\\
1380 & $3.0 \pm 0.6$\\
2378 & $1.3 \pm 0.3$\\
4850 & $1.2 \pm 0.2$\\
\end{tabular}
\end{table}

The response of the ATCA interferometer to the shortest spatial frequencies
is different from that of the Parkes and Molonglo radio telescopes. Hence,
it is difficult to make a direct comparison of the total flux densities.

\subsection{Polarisation}

Synchrotron radiation from electrons in a uniform magnetic field is
highly directional and so should be linearly polarised \citep{mof94a}.
Ideally, the orientation of the magnetic field may be determined from
the polarisation position angle. This assumes that the magnetic field
is an ordered field in a vacuum.

Relatively young SNRs are known to have quite large depolarisation
effects due to disorder in the magnetic field in the source itself
\citep{mof94a,mof94b}. Polarisation images of the $Q$ and $U$ Stokes
parameters were obtained which showed that there is no significant
polarisation in the extended source. This lack of polarisation is
confirmed by \citet{dun97}, who reports no apparent polarisation at
2.4 GHz.

\subsection{Distance and Size}

Distances to SNRs can occasionally be obtained indirectly either from
measuring optical velocities and proper motions or from positional
coincidences with H I and H II regions, OB associations, or pulsars.
However, the simple (but disputed) method of using the radio surface
brightness-to-diameter relationship $(\Sigma-D)$ \citep{cas96,cas98}
is adopted here to obtain a distance estimate for the observed source,
assuming that it is a SNR.

The radio surface brightness is defined as
\begin{equation}
\Sigma_\nu \equiv 1.505 \times 10^{-19} \, \frac{S_\nu}{\theta^2} \,
\mathrm{W \, m^{-2} \, Hz^{-1} \, sr^{-1}}
\end{equation}
where $S_\nu$ is in Janskys and $\theta$ is in arcminutes. The
${\Sigma-D}$ relation for Galactic shell remnants derived by \citet{cas98}
is given by
\begin{eqnarray}
\Sigma_{1 \, \mathrm{GHz}} & = & 2.07_{-1.24}^{+3.10} \times 10^{-17}
\nonumber\\
& & \times D^{(-2.38 \pm 0.26)} \, \mathrm{W \, m^{-2} \, Hz^{-1} \,
sr^{-1}}
\end{eqnarray}
where the diameter $D$ is in parsecs. The distance follows directly.

In the case of the observed object, the flux density $S_\nu$ at 1 GHz
was estimated (assuming the spectrum shown in Figure 6 peaks before
1 GHz) to be $3.5 \pm 0.7$ Jy. From equation (2), the radio surface
brightness was estimated to be $\Sigma_{1 \, \mathrm{GHz}} = 1.626
\times 10^{-21} \, \mathrm{W \, m^{-2} \, Hz^{-1} \, sr^{-1}}$. The
diameter of the object was then estimated from equation (3) to be
$D \approx 53 \, \mathrm{pc}$. This leads to an estimate of the
distance to the source of $\approx 10 \, \mathrm{kpc}$.

The uncertainty in these calculations are subject to the errors
inherent in the assumptions with the $\Sigma-D$ relation. These are
the assumptions that all shell remnants have the same radio luminosity
dependence on linear diameter, have the same supernova explosion
mechanism and energy, and are evolving into identical environments
\citep{cas98}. It has been concluded by \citet{gre84} that the
$\Sigma-D$ relation is severely limited and that the distance
determinations may have a factor of 3 error. However, by using known
distances to SNRs as distance ``calibrators'', \citet{cas98} find the
error in their derived $\Sigma-D$ relation to be $\sim 40\%$.

\section{Conclusions}

Mosaic observations of an extended source from the PMN survey were
carried out using the ATCA in early 1995 at frequencies of 1380 MHz and
2378~MHz. To complement these data, further observations of the source
were obtained from the MOST at a frequency 843~MHz.

The morphology of the extended object shows a distinct shell structure
which is optically thin. This is common to both SNRs and planetary
nebulae. In the case of a SNR, a shell structure would indicate that
the remnant was possibly produced by a Type I supernova resulting from
the collapse of a white dwarf star. Interactions with the interstellar
medium may have produced variations in the spherical nature of the
remnant's structure.

If the source is a SNR, the lack of significant polarisation suggests
that the intervening magnetic fields are disordered or that the SNR
candidate is at a large distance.

An estimate of the source's distance was made using the $\Sigma-D$
relation for Galactic shell remnants. This yielded a diameter of
$\approx 53 \, \mathrm{pc}$ and a distance of $\approx 10 \,
\mathrm{kpc}$. These results are consistent with the known
parameters of SNRs and that the apparent lack of linear polarisation is
likely due to the large distance. The distance estimate is of course
dependant on the source being a SNR.

Since interferometers do not respond to the shortest spatial
frequencies, it is difficult to compare total flux densities from ATCA
observations with those from the Parkes Radio Telescope. The 1.38 GHz
and 2.38 GHz results may be improved by using the Parkes telescope
which contains the lower spacings. Additional spectral information may
be obtained from further observations using the ATCA and the Parkes
telescope at X-Band (8000--9200 MHz). It would also be useful to obtain
X-ray spectra of the source to gain an estimate of the shock velocity
and thus determine the age and expansion rate of the SNR candidate.

\section*{Acknowledgements}

The authors would like to thank Robert Sault for providing assistance
with the data reduction in Miriad and Anne Green for her help in the
acquiring and reduction of the Molonglo radio map and for her useful
comments on the manuscript. Thanks also are due to Simon Ellingsen and
Richard Dodson for their useful comments.

\end{document}